\documentclass[prd,twocolumn,nofootinbib,superscriptaddress,amssymb,amsfonts,amsmath,preprintnumbers]{revtex4-2}
\usepackage{graphicx,slashed,xcolor,multirow,bbold,mathtools,sidecap,tikz,bm,ulem,enumitem,booktabs,array}
\usepackage[colorlinks=true,linktoc=page,linkcolor=purple,citecolor=teal,urlcolor=magenta]{hyperref}
\usepackage[compat=1.1.0]{tikz-feynman}
\graphicspath{ {F95/} }
\usepackage{siunitx,adjustbox}

\definecolor{lime}{HTML}{A6CE39}
\DeclareRobustCommand{\orcidicon}{\hspace{-2.1mm}
\begin{tikzpicture}
\draw[lime,fill=lime] (0,0.0) circle [radius=0.13] node[white] {{\fontfamily{qag}\selectfont \tiny ID}}; \draw[white,fill=white] (-0.0525,0.095) circle [radius=0.007]; 
\end{tikzpicture} \hspace{-3.7mm} }
\foreach \x in {A, ..., Z} {\expandafter\xdef\csname orcid\x\endcsname{\noexpand\href{https://orcid.org/\csname orcidauthor\x\endcsname} {\noexpand\orcidicon}}}

\let\emph\textit

\begin{document}

\preprint{PSI-PR-23-20, ZU-TH 28/23, ICPP-71}

\title{$SU(2)_L$ triplet scalar as the origin of the $95\,$GeV excess?}

\author{Saiyad Ashanujjaman\orcidA{}}
\email{saiyad.a@iopb.res.in}
\affiliation{Department of Physics, SGTB Khalsa College, Delhi 110007, India}
\affiliation{Department of Physics and Astrophysics, University of Delhi, Delhi 110007, India}

\author{Sumit Banik\orcidB{}}
\email{sumit.banik@psi.ch}
\affiliation{Physik-Institut, Universität Zürich, Winterthurerstrasse 190, CH–8057 Zürich, Switzerland}
\affiliation{Paul Scherrer Institut, CH–5232 Villigen PSI, Switzerland}

\author{Guglielmo Coloretti\orcidC{}}
\email{guglielmo.coloretti@physik.uzh.ch}
\affiliation{Physik-Institut, Universität Zürich, Winterthurerstrasse 190, CH–8057 Zürich, Switzerland}
\affiliation{Paul Scherrer Institut, CH–5232 Villigen PSI, Switzerland}

\author{Andreas Crivellin\orcidD{}}
\email{andreas.crivellin@cern.ch}
\affiliation{Physik-Institut, Universität Zürich, Winterthurerstrasse 190, CH–8057 Zürich, Switzerland}
\affiliation{Paul Scherrer Institut, CH–5232 Villigen PSI, Switzerland}

\author{Bruce Mellado}
\email{bmellado@mail.cern.ch}
\affiliation{School of Physics and Institute for Collider Particle Physics, University of the Witwatersrand, Johannesburg, Wits 2050, South Africa}
\affiliation{iThemba LABS, National Research Foundation, PO Box 722, Somerset West 7129, South Africa}

\author{Anza-Tshilidzi Mulaudzi}
\email{anza-tshilidzi.mulaudzi@cern.ch}
\affiliation{School of Physics and Institute for Collider Particle Physics, University of the Witwatersrand, Johannesburg, Wits 2050, South Africa}
\affiliation{iThemba LABS, National Research Foundation, PO Box 722, Somerset West 7129, South Africa}

\begin{abstract}
We explore the possibility that an $SU(2)_L$ triplet scalar with hypercharge $Y=0$ is the origin of the $95\,$GeV diphoton excess. For a small mixing angle with the Standard Model Higgs, its neutral component has naturally a sizable branching ratio to $\gamma\gamma$ such that its Drell-Yan production via $pp\to W^*\to H H^\pm$ is sufficient to obtain the desired signal strength, where $H^\pm$ is the charged Higgs component of the triplet. The predictions of this setup are: 1) The $\gamma\gamma$ signal has a $p_T$ spectrum different from gluon fusion but similar to associated production. 2) Photons are produced in association with tau leptons and jets, but generally do not fall into the vector-boson fusion category. 3) The existence of a charged Higgs with $m_{H^\pm}\approx\!(95\pm5)\,$GeV leading to $\sigma(pp\to \tau\tau\nu\nu)\approx0.4\,$pb, which is of the same level as the current limit and can be discovered with Run 3 data. 4) A positive definite shift in the $W$ mass as suggested by the current global electroweak fit.
\end{abstract}
\maketitle

\section{Introduction} 
The Standard Model (SM) is the currently accepted theoretical description of the known constituents and interaction of matter. It has been successfully tested in precision experiments~\cite{ALEPH:2005ab,HeavyFlavorAveragingGroup:2022wzx,ParticleDataGroup:2022pth} and the Brout-Englert-Higgs boson~\cite{Higgs:1964ia,Englert:1964et,Higgs:1964pj,Guralnik:1964eu}, the last missing piece, was finally discovered in 2012 at CERN~\cite{Aad:2012tfa,Chatrchyan:2012ufa,CDF:2012laj}. In fact, this $125\,$GeV particle has properties consistent with the ones predicted by the SM~\cite{Chatrchyan:2012jja,Aad:2013xqa,ATLAS:2016neq,Langford:2021osp,ATLAS:2021vrm}. However, this does not exclude the existence of additional scalar bosons, as long as their role in electroweak symmetry breaking is subleading and their production cross sections are smaller than the ones of the SM-like Higgs~\cite{Kole:2022pqz,Palmer:2021gmo}. 

The minimality of the SM Higgs sector, {\it i.e.}~the existence of a single $SU(2)_L$ doublet scalar that simultaneously gives mass to the electroweak (EW) gauge bosons and all fermions, is not guaranteed by any theoretical principle or symmetry. A plethora of such extensions have been proposed in the literature, including the addition of $SU(2)_L$ singlets~\cite{Silveira:1985rk,Pietroni:1992in,McDonald:1993ex}, doublets~\cite{Lee:1973iz,Haber:1984rc,Kim:1986ax,Peccei:1977hh,Turok:1990zg} and triplets~\cite{Konetschny:1977bn,Cheng:1980qt,Lazarides:1980nt,Schechter:1980gr,Magg:1980ut,Mohapatra:1980yp}.

While Large Hadron Collider (LHC) searches for new particles did not lead to any discovery (yet), there are interesting hints for new scalar bosons~\cite{Fischer:2021sqw}. In particular, CMS~\cite{CMS:2018cyk,CMS:2023yay,CMS:2022goy} searches hint toward a neutral scalar $H$ decaying into two photons at 95$\,$GeV. This is compatible with the latest ATLAS result~\cite{ATLAS:2023jzc} and supported by $Z$-strahlung with $H\to b\bar b$ at LEP~\cite{LEPWorkingGroupforHiggsbosonsearches:2003ing}, as well as by $\tau\tau$~\cite{CMS:2022goy} and $WW$~\cite{CMS:2022uhn,ATLAS:2022ooq,Coloretti:2023wng} searches. In fact, combining these channels results in a global significance of $3.8\sigma$~\cite{Bhattacharya:2023lmu}.

So far, explanations of the $95\,$GeV excesses in terms of $SU(2)_L$ singlets and/or $SU(2)_L$ doublets were proposed in the literature~\cite{Cao:2016uwt,Crivellin:2017upt,Haisch:2017gql,Fox:2017uwr,Liu:2018xsw,Wang:2018vxp,Liu:2018ryo,Biekotter:2019kde,Cline:2019okt,Choi:2019yrv,Kundu:2019nqo,Cao:2019ofo,Biekotter:2020cjs,Abdelalim:2020xfk,Heinemeyer:2021msz,Biekotter:2022jyr,Iguro:2022dok,Li:2022etb,Iguro:2022fel,Biekotter:2023jld,Bonilla:2023wok,Azevedo:2023zkg,Banik:2023ecr,Biekotter:2023oen,Escribano:2023hxj,Belyaev:2023xnv}, which all respect custodial symmetry at tree-level. For higher {dimensional} $SU(2)_L$ representations, the measurement of the $\rho$-parameter restricts the vacuum expectation value (VEV) of the new scalar to be $\lesssim \mathcal{O}(1)\,$GeV~\cite{ParticleDataGroup:2022pth} and except for the $SU(2)_L$ triplet with hypercharge $Y=0$ multiply charged scalars at the same mass scale are unavoidable which is problematic with respect to LHC searches~\cite{D0:2011eug,CDF:2011wxl,ATLAS:2014kca,CMS:2016cpz}.\footnote{{For small mass-splitting among the $SU(2)_L$ components, LHC searches for multiply charged scalars would exclude scenarios with a neutral Higgs with a mass around $\sim 95\,$GeV~\cite{Heeck:2022fvl,Ashanujjaman:2022ofg}. However, nondegenerate scenarios, with the heavier multiply charged Higgses decaying into (off-shell) neutral Higgses and $W$-bosons, could still be consistent with the LHC searches~\cite{Chakrabarti:1998qy,Akeroyd:2005gt,Aoki:2011pz,Ashanujjaman:2021txz}. The phenomenology of such mass spectra has been studied in Refs.~\cite{Ashanujjaman:2022tdn,Ashanujjaman:2023tlj}.}} It is well known that this field provides a positive definite shift in the $W$ mass (with respect to~the SM prediction)~\cite{Chabab:2018ert,FileviezPerez:2022lxp, Cheng:2022hbo,Chen:2022ocr,Rizzo:2022jti,Chao:2022blc,Wang:2022dte,Shimizu:2023rvi,Lazarides:2022spe,Senjanovic:2022zwy,Crivellin:2023gtf,Chen:2023ins}, as motivated by the current global electroweak fit~\cite{deBlas:2022hdk,Athron:2022isz,Bagnaschi:2022whn} (driven by the CDF~II result~\cite{CDF:2022hxs}). However,  its collider phenomenology has been barely studied. In this article, we study the viability of $Y=0$ triplet as an alternative in addressing the hints for a $\approx\!95\,$GeV scalar.

\section{Phenomenology}

The SM extended with an $SU(2)_L$ triplet scalar with hypercharge 0, is commonly referred to as the $\Delta$SM~\cite{Ross:1975fq,Gunion:1989ci,Chankowski:2006hs,Blank:1997qa,Forshaw:2003kh,Chen:2006pb,Chivukula:2007koj,Bandyopadhyay:2020otm}. It contains an additional charged scalar  $H^\pm$ and a neutral one $H$ which acquires a vacuum expectation value $v_\Delta$ in the process of spontaneous symmetry breaking. Importantly, without mixing $H$ couples only to $W$ bosons at tree-level, while the $CP$-even mixing angle $\alpha$ induces couplings to SM fermions. Furthermore, charged Higgs loops modify both $h\to\gamma\gamma$ and $H\to\gamma\gamma$. A detailed description of the model is provided in the Appendix.

\subsection{Perturbative unitarity and vacuum stability}

The $\Delta$SM parameter space can be constrained by vacuum stability and perturbative unitarity. The region between the red lines in Fig.~\ref{mainplot} is allowed by both criteria and the explicit calculation of the constraints is given in the Appendix.

\begin{figure*}
\centering
\hspace{-0.5cm}
\includegraphics[scale=0.4]{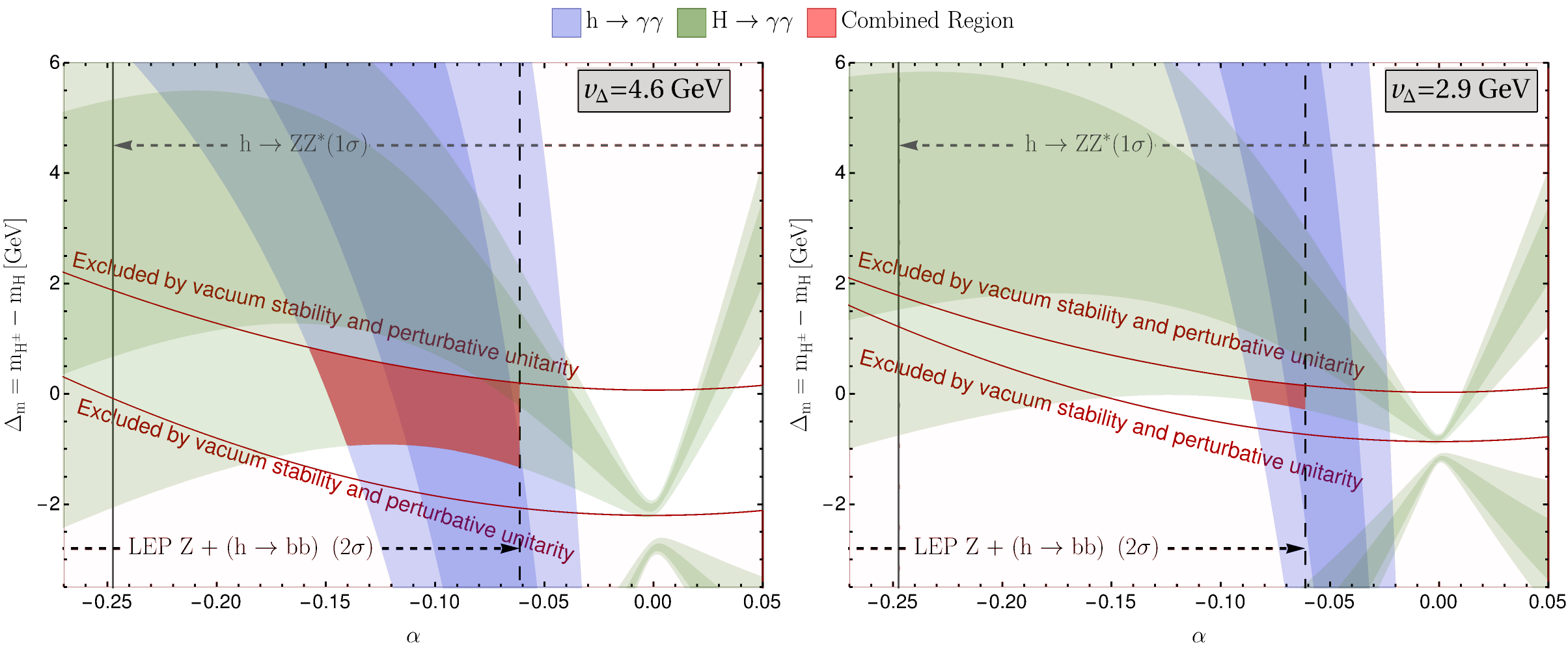}
\caption{Preferred regions ($1\sigma$ and $2\sigma$) by the $h\to\gamma\gamma$ signal strength (blue) and the 95\,GeV $H\to\gamma\gamma$ excess (green) in the $\alpha-\Delta_m$ plane for the two values of $v_\Delta$ corresponding to the two $m_W$ benchmark points. The region between the two red lines is allowed by vacuum stability and perturbative unitarity. The dashed vertical line indicates the region preferred by the LEP measurement of $Z+(H\to bb)$, and the region to the right of the solid vertical line is preferred by the $h\to ZZ^*$ signal strength at $1\sigma$ level.}
\label{mainplot}
\end{figure*}

\subsection{$W$ mass}
The latest ATLAS update of $m_W = 80.360(16)$~\cite{ATLAS:2023fsi} (superseding the 2017 result~\cite{ATLAS:2017rzl}) as well as the LHCb result $m_W = 80.354(32)$~\cite{LHCb:2021bjt} are significantly smaller compared to $m_W = 80.4335(94)\,$GeV obtained by CDF~II. When combined with D0~\cite{CDF:2022hxs} and LEP~\cite{ALEPH:2013dgf}, this lead to a naive global average of $m_W = 80.406(7)\,$GeV. Because there is considerable tension between these measurements ($\chi^2/{\rm dof} = 4.3$), we inflate the error on $m_W$ to 0.015$\,$GeV to get a conservative average of~\cite{Crivellin:2023gtf}\footnote{This {\it naive} average agrees well with the one obtained in a sophisticated fit performed by {\tt HEPfit}~\cite{deBlas:2022hdk} prior to the ATLAS update.}
\begin{align}
m_W^{\rm comb} = (80.406 \pm 0.015){\rm\,GeV}.
\end{align}
Comparing this with the SM prediction of $m_W^{\rm SM} = 80.3499(56)\,$GeV~\cite{deBlas:2022hdk,Sirlin:1983ys,Djouadi:1987gn,Avdeev:1994db,Chetyrkin:1995js,Chetyrkin:1995ix,Awramik:2003rn,Degrassi:2014sxa,ParticleDataGroup:2022pth}, with $m_t = 172.5(0.7)\,$GeV~\cite{ParticleDataGroup:2022pth}, the discrepancy of $56\,$MeV amounts to $3.7\sigma$. If we disregarded the CDF~II result, we find an average of 
\begin{align}
m_W^{\rm comb~(w/o~CDF~II)} = (80.372 \pm 0.010){\rm\,GeV},
\end{align}
which corresponds to a discrepancy of 22$\,$MeV ($2.2\sigma$). 

In the $\Delta$SM, we have
\begin{align}
m_{W}^2 &= \frac{g^2}{4} (v^2 + 4 v_\Delta^2), \hspace{7mm}  m_Z^2 = \frac{g^2}{4 \cos \theta_W^2} v^2.
\label{eq:Wmass}
\end{align}
Therefore, $v_\Delta$ of a few GeV can easily alter the $m_W$ prediction in the desired direction. As such, $m_W^{\rm comb}$ requires $v_\Delta = 4.60 ^{+0.58}_{-0.66}\,$GeV, while $m_W^{\rm comb~(w/o~CDF~II)}$ requires $v_\Delta = 2.89 ^{+0.59}_{-0.75}\,$GeV. 

\subsection{SM Higgs signal strength}
Through the quartic interactions $H^\pm$ contributes to the diphoton decay rate of the SM Higgs $h$ (see Fig.~\ref{fig:ppToZgamma} left). The corresponding signal strength, with respect to~the SM one, is given by
\begin{align}
& \mu_{h,\gamma\gamma} = {\Gamma_{h \to \gamma\gamma}}/{\Gamma_{h \to \gamma\gamma}^{\rm SM}} = |\kappa_\gamma^2|\,,
\end{align}
with
\begin{align}
\kappa_{\gamma} \approx \cos\alpha +\dfrac{A_{h H^\pm H^\mp} v}{2m_{H^\pm}^2} \dfrac{\beta^0_H\left(\frac{4 m_{H^\pm}^2}{m_h^2}\right)}{\dfrac{4}{3} \beta^{1/2}_H\left(\frac{4 m_t^2}{m_h^2}\right) + \beta^1_H\left(\frac{4 m_W^2}{m_h^2}\right)},
\end{align}
and the loop functions~\cite{Gunion:1989we} are given in Appendix.

Combining the most recent measurements of CMS~\cite{CMS:2021kom} and ATLAS~\cite{ATLAS:2022tnm}, $\mu_{h, \gamma\gamma}^{\rm CMS} = 1.12^{+0.09}_{-0.09}$ and  $\mu_{h, \gamma\gamma}^{\rm ATLAS} = 1.04^{+0.10}_{-0.09}$, respectively, we get the weighted average
\begin{align}
\mu_{h,\gamma\gamma}^{\rm exp} = 1.08^{+0.07}_{-0.06}.
\end{align}
The resulting preferred regions at the $1\sigma$ and $2\sigma$ level are shown in blue in Fig.~\ref{mainplot}.

While the $h\to\gamma\gamma$ signal strength is the most precise measured one, it is affected by $h$-$H$ mixing and the $H^\pm$-loop contribution so that cancellations occur. Therefore, the second-best measured SM Higgs signal $h\to ZZ^*$~\cite{CMS:2022dwd,ATLAS:2020rej} provides a complementary constraint of~\cite{ParticleDataGroup:2022pth}
\begin{equation}
    \mu_{h,ZZ^*}^{\rm exp}=1.02\pm 0.08\,,
\end{equation}
which, to a very good approximation, is only sensitive to the mixing angle $\alpha$. The region on the right of the solid vertical line in Fig.~\ref{mainplot} is compatible with $\mu_{h,ZZ^*}^{\rm exp}$ at the $1\sigma$ level.

\subsection{diphoton excess}

While nearly all relevant decay modes of $H$ can be obtained from a rescaling of the widths of a SM-like Higgs with a mass of 95\,GeV by multiplying with $\sin^2\alpha$, the decay $H\to WW^*$ is already generated at tree-level via $v_\Delta$ and $H\to\gamma\gamma$ receives loop contributions from the charged Higgs as well as from $W$ loops:\footnote{Only $Z\gamma$ also receives an additional direct contribution from the $W$ loop, which is already present for $\sin\alpha=0$, but the corresponding branching ratio is negligibly small.}  
\begin{widetext}
\begin{align}
\Gamma(H\to \gamma\gamma) \approx \dfrac{\alpha_{\rm em}^2 g_2^2 m_{H}^3}{1024\pi^3m_W^2} \left|- \dfrac{4}{3} \sin\alpha \, \beta_H^{1/2}\left(\frac{4 m^2_t}{m^2_H}\right) + \left( -\sin\alpha + \dfrac{4 v_\Delta}{v} \cos\alpha\right) \beta_H^1\left(\frac{4 m^2_W}{m^2_H}\right) + \dfrac{A_{H H^\pm H^\mp} v}{2m_{H^{\pm}}^2} \beta_H^0\left(\frac{4 m^2_{H^{\pm}}}{m^2_H}\right) \right|^2.
\end{align}
\end{widetext}
Here, $\alpha_{\rm em}$ at $q^2 = 0$ numerically approximates well the NLO QED corrections. 

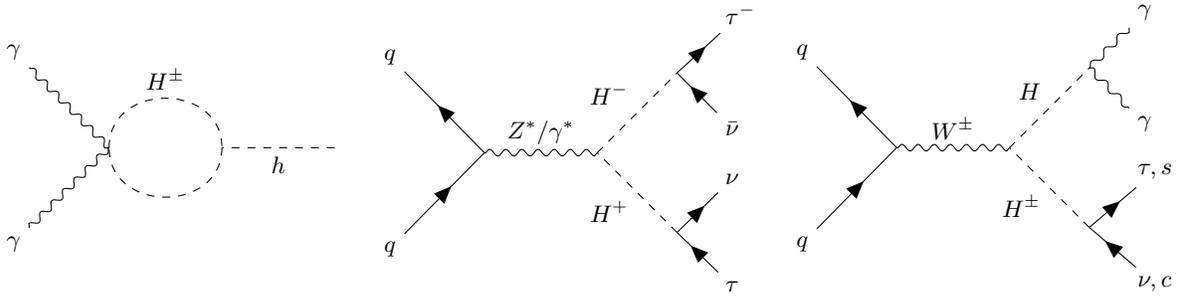
\begin{figure*}[t!]
    \centering
    \begin{tikzpicture}[baseline=(current bounding box.center)]
        \begin{feynman}
            \vertex (a);
            \vertex [above left=1.5cm of a] (b) {$\gamma$};
            \vertex [below left=1.5cm of a] (c) {$\gamma$};
            \vertex [right=1.5cm of a] (d) ;
            \vertex [right=1.5cm of d] (e);
            \diagram{
                (b) -- [boson] (a) -- [boson] (c);
                (a) -- [scalar, half left, edge label=$H^\pm$] (d);
                (d) -- [scalar, half left] (a);
                (e) -- [scalar, edge label = $h$] (d)
            };
        \end{feynman}
    \end{tikzpicture}
    ~~~
    \begin{tikzpicture}[baseline=(current bounding box.center)]
        \begin{feynman}
            \vertex (a);
            \vertex [above left=1.5cm of a] (c) {$q$};
            \vertex [below left=1.5cm of a] (d) {$q$};
            \vertex [right=1.5cm of a] (b) ;
            \vertex [above right=1.5cm of b] (e);
            \vertex [below right=1.5cm of b] (f);
            
            \vertex [above right=0.75cm of e] (i) {$\tau^-$};
            \vertex [below right=0.75cm of e] (j) {$\Bar{\nu}$};
            \vertex [above right=0.75cm of f] (k) {$\nu$};
            \vertex [below right=0.75cm of f] (l) {$\tau$};
            \diagram{
                (d) -- [fermion] (a) -- [fermion] (c);
                (a) -- [boson, edge label=$Z^{*}/\gamma^{*}$] (b);
                (f) -- [scalar, edge label=$H^+$] (b) -- [scalar, edge label=$H^-$] (e);
                (j) -- [fermion] (e) -- [fermion] (i);
                (l) -- [fermion] (f) -- [fermion] (k);
            };
        \end{feynman}
    \end{tikzpicture}
    ~~
    \begin{tikzpicture}[baseline=(current bounding box.center)]
        \begin{feynman}
            \vertex (a);
            \vertex [above left=1.5cm of a] (c) {$q$};
            \vertex [below left=1.5cm of a] (d) {$q$};
            \vertex [right=1.5cm of a] (b) ;
            \vertex [above right=1.5cm of b] (e);
            \vertex [below right=1.5cm of b] (f);
            
            \vertex [above right=0.75cm of e] (i) {$\gamma$};
            \vertex [below right=0.75cm of e] (j) {$\gamma$};
            \vertex [above right=0.75cm of f] (k) {$\tau,s$};
            \vertex [below right=0.75cm of f] (l) {$\nu,c$};
            \diagram{
                (d) -- [fermion] (a) -- [fermion] (c);
                (a) -- [boson, edge label=$W^\pm$] (b);
                (f) -- [scalar, edge label=$H^\pm$] (b) -- [scalar, edge label=$H$] (e);
                (j) -- [boson] (e) -- [boson] (i);
                (l) -- [fermion] (f) -- [fermion] (k);
            };
        \end{feynman}
    \end{tikzpicture}
    \caption{Feynman diagrams showing the modification of $h\to \gamma\gamma$ (left), the DY processes $pp\to Z^*,\gamma^*\to H^+H^-\to \tau^+\tau^-\nu\bar \nu$ (middle) and $pp\to W^*\to H^{\pm}H^{0}$ (right).}
    \label{fig:ppToZgamma}
\end{figure*}

For a small mixing angle $\alpha$, $H$ is mainly produced via the Drell-Yan (DY) process $pp\to W^*\to H^\pm H$ (see Fig.~\ref{fig:ppToZgamma} right) with a leading order (LO) cross section of 1.77$\,$pb for $m_{H^\pm} \approx m_H = 95\,$GeV. While the QCD corrections have not been estimated so far for the $\Delta$SM, it is obvious that they pertain dominantly to the hadronic ends of the processes and are thus expected to be the same as for sleptons or $SU(2)_L$ triplet leptons. The latter has been calculated in Ref.~\cite{Ajjath:2023ugn}, resulting in a correction factor of 1.15, by which we naively rescale the LO cross section (computed with {\tt MadGraph5aMC@NLO}~\cite{Alwall:2011uj}) to obtain $\approx2\,$pb. In addition, $H$ is also produced via gluon-gluon fusion (ggF) and vector boson fusion (VBF) processes through the mixing with $h$. The corresponding cross section is calculated by multiplying the production cross section of a SM-like 95$\,$GeV Higgs by $\alpha^2$. Neglecting the subdominant contribution from VBF, and using $\sigma[pp \to h(95)]\approx68\,$pb~\cite{LHCHiggsCrossSectionWorkingGroup:2016ypw,Graudenz:1992pv,Spira:1995rr,Anastasiou:2002yz,Harlander:2002wh,Harlander:2001is,Aglietti:2006tp,Li:2015kaa,Anastasiou:2006hc,Harlander:2005rq,Ravindran:2003um}, we thus have
\begin{align}
\sigma[pp \to H \to \gamma\gamma] \approx {\rm Br}[H\to\gamma\gamma] \times \left(2 + 68 \alpha^2 \right) {\rm pb}.
\label{eq:CMS_diphoton}
\end{align}
Normalizing the signal strength to the one of a hypothetical SM-like Higgs with the same mass~\cite{LHCHiggsCrossSectionWorkingGroup:2016ypw}, we find numerically 
\begin{align}
\mu_{H,\gamma \gamma} \approx (21.5 + 719\alpha^2) \times {\rm Br}[H \rightarrow \gamma \gamma].
\end{align}
This has to be compared to the combination of the CMS and ATLAS analyses of a low mass $\gamma\gamma$ searches of~\cite{Biekotter:2023oen}\footnote{{Note that the signal strength of $H$ is normalized with respect to~an SM-like Higgs with the same mass. While the latter is mainly produced via ggF and VBF processes, the former is dominantly produced via the DY process $pp \to W^* \to H^\pm H$ while the other production modes are too a good approximation only induced via the mixing with $h$. Note that, while in the limit of zero mixing between the SM Higgs and the triplet Higgs, $H$ is fermiophobic, this region in parameter space is, contrary to the setup of Ref.~\cite{Delgado:2016arn}, not excluded due to the charged Higgs contribution to $H\to\gamma\gamma$. Furthermore, for $\alpha \neq 0$, couplings to fermions are induced.}}
\begin{equation}
\mu_{H,\gamma \gamma}^{\rm exp} = 0.27^{+0.10}_{-0.09}\,.
\end{equation}
The resulting preferred regions are shown in green in Fig.~\ref{mainplot}.\footnote{{Note that our model has similarities with one of the ``square'' benchmark scenarios of Ref.~\cite{Haisch:2017gql}, where the 95$\,$GeV excess was studied in the context of the type-I two-Higgs-doublet model. There, in the fermiophobic limit, $pp \to W^{\pm*} \to H^\pm H$ is the dominant production mode. However, the model in Ref.~\cite{Haisch:2017gql} predicts an additional pseudoscalar with $\approx 80\,$GeV while the Higgs potential allows for more freedom than our setup.}}.

\subsection{$Zbb$, $WW$ and $\tau\tau$}

While Br$[H\to WW]$ is large for a very small mixing angle $\alpha$, the resulting effect in $\gamma\gamma$ would be too high if one aims at the central value of the cross section of Ref.~\cite{Coloretti:2023wng}. Therefore, $\alpha$ cannot be too small, and it is possible to explain the $Zbb$ excess of LEP which requires
\begin{equation}
\mu_{b b}^{\exp }=\frac{\sigma^{\exp }\left(e^{+} e^{-} \rightarrow ZH\right)}{\sigma^{\mathrm{SM}}\left(e^{+} e^{-} \rightarrow Z H\right)} {\rm Br}\left(H \to b \bar{b}\right) =0.117 \pm 0.057.    
\end{equation}
For tau decays, the central values of the signal strength $\mu_{\tau\tau}^{\rm exp}=1.2\pm 0.5$ cannot be fully explained, which is a general feature of most SM extensions addressing the $95\,$GeV excess~\cite{Iguro:2022dok}, the error is too large to draw a conclusion here.
\subsection{$pp\to H^+H^-\to \tau^+\tau^-\nu\bar \nu$}

The charged Higgs in general dominantly decays to $\tau\nu$. Therefore, its pair production and subsequent decays, {\it i.e.}~$pp\to Z^*,\gamma^* \to H^+H^-\to \tau^+\tau^-\nu\bar\nu$ (see Fig.~\ref{fig:ppToZgamma} middle), leads to a collider signature searched for in the context of supersymmetric tau partners~\cite{ATLAS:2019gti,CMS:2019eln,CMS:2022rqk,ATLAS:2023djh}. While CMS~\cite{CMS:2022rqk} provides an upper bound on the cross section and observes a weaker limit than expected, ATLAS~\cite{ATLAS:2023djh} observes a stronger limit than expected but does not provide a bound on the total cross section. Since both bounds deviate from the expected limit by $\approx1\,\sigma$ level, but in opposite directions, we will thus use the expected limit on the cross section provided by CMS~\cite{CMS:2022rqk} of $0.34^{+0.24}_{-0.12}\,$pb. Using once more {\tt MadGraph5aMC@NLO} at LO, we find a production cross section of $0.86\,$pb which we again multiply by a factor $1.15$~\cite{Ruiz:2015zca,Ajjath:2023ugn} to include NLO QCD effects. Taking into account that CMS and ATLAS assume a 100\% branching ratio of the stau to tau and neutralino, while we have Br$[H^\pm\to \tau^\pm \nu_\tau ]\approx 0.66\pm 0.03$~\cite{LHCHiggsCrossSectionWorkingGroup:2016ypw,Braaten:1980yq, Sakai:1980fa,Inami:1980qp,Gorishnii:1983cu,Gorishnii:1990kd, Gorishnii:1990zu,Gorishnii:1991zr,Djouadi:1997rj,Degrassi:2005mc, Passarino:2007fp,Actis:2008ts,Djouadi:1990aj,Chetyrkin:1996sr,Baikov:2005rw,Spira:1991tj}\footnote{Since in our case the branching ratio is dominated by $\tau\nu$ and $cs$, the error on Br$[H^\pm\to cs]$ is dominating the error of Br$[H^\pm\to \tau^\pm \nu_\tau ]$.}, a cross section of $\approx 0.44\pm0.03\,$pb is predicted. This is in slight tension with the 95\% exclusion limit. 

Let us therefore consider the option to reduce Br$[H^\pm\to \tau\nu]$ by increasing the mass splitting $\Delta m$ such that Br$[H^\pm \to H W^{*}]$ becomes sizable:\footnote{Note that Br$[H^\pm \to H^* W]$ is much smaller such that it can be neglected.} 
\begin{align}
&\Gamma(H^\pm \to H W^{*}) = \frac{9g^4m_{H^\pm}}{512\pi^3} \lambda_{HH^\pm W}^2 G\left(\frac{m_H^2}{m_{H^\pm}^2},\frac{m_W^2}{m_{H^\pm}^2}\right),
\end{align}
where $\lambda_{HH^\pm W} = 2\cos\alpha \cos\beta - \sin\alpha \sin\beta$, and the loop function $G(x,y)$ is given in the Appendix.

As one can see in Sec.~3 in the Appendix, choosing $v_\Delta= 0.86\,$GeV as a benchmark point, allows for a small region in parameter space with sizable mass splitting, that is allowed by the vacuum stability and perturbative unitarity\footnote{Note that, for sizeable $\alpha$, the requirements of vacuum stability and perturbative unitarity dictate that $\Delta_m \approx 22\alpha - 3.75$ GeV.} as well as compatible with $h\to\gamma\gamma,ZZ^*$, $H\to\gamma\gamma$ and $Zbb$. Note that this scenario predicts a small positive shift in the $W$ mass.

\section{Conclusions and Outlook}

\begin{figure}[t!]
\centering
\hspace*{-4mm}
\includegraphics[scale=0.6]{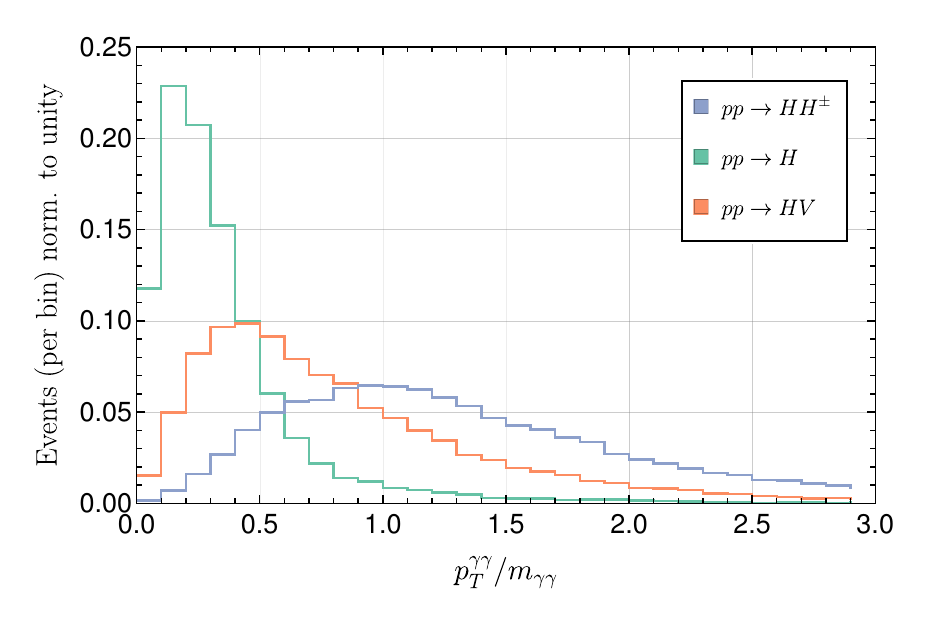}
\caption{Transverse momentum normalized to the invariant mass of the photon pair system for different production mechanisms of a 95$\,$GeV scalar $H$: VH (orange), ggF (green), DY production in the triplet model (blue).}
\label{fig:pt12sim}
\end{figure}

In summary, the predictions if the neutral component of the $SU(2)_L$ triplet with hypercharge 0 is the origin of the 95\,GeV excess are:
\begin{itemize}
    \item LHC Run 3 shows a stau-like excess.
    \item Positive shift in the $W$ mass.
    \item $H$ is produced in association with jets and $\tau$ leptons.
    \item A charged Higgs with a mass below $\approx\!100\,$GeV which could be very well studied at future $e^+e^-$ colliders~\cite{FCC:2018evy,Aicheler:2012bya,Behnke:2013lya,An:2018dwb}.
    \item A significantly broader $p_T$ spectrum of the diphoton system compared to ggF, as shown in Fig.~\ref{fig:pt12sim}.\footnote{While this information is currently not available, it can be used in future analyses as a discriminator. To compare the $p_T$ of the diphoton system of the $\Delta$SM to the SM, we generated 100k events at NLO using {\tt MadGraph5aMC@NLO} with the parton shower performed by {\tt Pythia8.3}~\cite{Sjostrand:2014zea} and the detector simulation for the CMS detector~\cite{CMS:2023yay}, carried out with {\tt Delphes}~\cite{deFavereau:2013fsa}. The UFO model file at NLO of the $\Delta$SM was built using {\tt FeynRules}~\cite{Degrande:2011ua,Alloul:2013bka, Degrande:2014vpa} and to increase the efficiency of the simulation, the decay of $H$ to a  photon pair was forced using {\tt MadSpin}~\cite{BuarqueFranzosi:2019boy}.}  
\end{itemize}

\begin{acknowledgments}
S.B., G.C.~and A.C.~thank Michael Spira for useful discussions. S.A.~acknowledges the support via the Core Research Grant CRG/2018/004889 from SERB, Government of India. The work of A.C.~is supported by a professorship grant from the Swiss National Science Foundation (No.\ PP00P21\_76884). B.M.~gratefully acknowledges the South African Department of Science and Innovation through the SA-CERN program, the National Research Foundation, and the Research Office of the University of the Witwatersrand for various forms of support. S.A.~thanks Rameswar Sahu for useful discussions.
\end{acknowledgments}

\appendix

\maketitle


\section{The Model}

The SM supplemented by a $SU(2)_L$ triplet scalar with hypercharge 0, is commonly referred to as the $\Delta$SM~\cite{Ross:1975fq,Gunion:1989ci,Chankowski:2006hs,Blank:1997qa,Forshaw:2003kh,Chen:2006pb,Chivukula:2007koj,Bandyopadhyay:2020otm}. The scalar sector consists of the SM doublet $\Phi$, and the triplet $\Delta$:
\begin{equation}
\Phi = \begin{pmatrix} \phi^+ \\ \phi^0 \end{pmatrix}, \quad \Delta = \frac{1}{2} \begin{pmatrix} \delta^0 & \sqrt{2}\delta^+ \\ \sqrt{2}\delta^- & -\delta^0 \end{pmatrix},
\end{equation}
with $\delta^0$ being real and $\delta^- = (\delta^+)^*$.
The covariant derivative for $SU(2)_L$ is fixed, in the usual conventions, by the generators, {\it i.e.}~$T_{k} = \frac{\sigma_{k}}{2}$ for the doublet, with $\sigma_{k}$ being the Pauli matrices. This then fixes the structure constants $f_{ijk} = i \epsilon_{ijk}$ and the covariant derivative in the adjoint representation for the triplet is
\begin{equation}
    D_{\mu} \Delta = \partial_{\mu} \Delta  - i g_2 \left[ \frac{\sigma_{k}}{2} \; W_{\mu}^{k}, \; \Delta \right]\,,
\end{equation}
where the square bracket stands for the commutator. Note that therefore the $SU(2)_L$ gauge boson interactions with the triplet are a factor of 2 higher than for a doublet.

Since the triplet cannot have direct couplings to quarks or leptons, the scalar potential 

\begin{align}
\label{eq:pot}
V & = -\mu_\Phi^2 \Phi^\dag \Phi + \frac{\lambda_\Phi}{4} \left(\Phi^\dag \Phi\right)^2 - \mu_\Delta^2 {\rm Tr}\left(\Delta^\dag \Delta\right) \\ \nonumber &+ \frac{\lambda_\Delta}{4} \left[ {\rm Tr}\left(\Delta^\dag \Delta\right) \right]^2
 + \mu \Phi^\dag \Delta \Phi + \lambda_{\Phi \Delta} \Phi^\dag \Phi {\rm Tr}\left(\Delta^\dag \Delta\right),
\end{align}
describes its remaining interactions.
Note that Eq.~\eqref{eq:pot} has a global $O(4)_H \times O(3)_\Delta$ symmetry softly broken by the $\mu$-term. After  electroweak symmetry breaking, $\phi^0$ and $\delta^0$ acquire the VEVs $\langle \sqrt{2} \phi^0 \rangle = v \approx 246\,$GeV and $\langle \delta^0 \rangle = v_\Delta$. The minimization conditions
\begin{align}
\begin{aligned}
\mu_\Phi^2 &= - \mu \frac{v_\Delta}{2}+\frac{1}{4} v^2 \lambda _{\Phi }+\frac{1}{2} \lambda _{\Phi \Delta } v_\Delta^2,
\label{eq:minimization-1}
\\
\mu_{\Delta}^2 &=  - \mu \frac{v^2}{4 v_\Delta}+\frac{1}{2} v^2 \lambda _{\Phi \Delta }+\frac{1}{4} \lambda _{\Delta } v_\Delta^2,
\end{aligned}
\end{align}
can then be used to eliminate $\mu_\Phi^2$ and $\mu_\Delta^2$ in terms of the other parameters of Eq.~\eqref{eq:pot}. The scalar mass matrices, in the bases ($\phi^+, \delta^+$) and (Re$(\phi^0)$, $\delta^0$), are
\begin{align}
\begin{aligned}
M^2_\pm &= \mu \begin{pmatrix}
v_\Delta & \frac{v}{2} \\[.1cm]
\frac{v}{2} & \frac{v^2}{4 v_\Delta}
\end{pmatrix}\,,
\\M^2_0 &= \begin{pmatrix}
\lambda_\Phi  \frac{v^2}{2} & \lambda_{\Phi \Delta } v v_\Delta- \mu \frac{v}{2} \\[.1cm]
\lambda_{\Phi \Delta } v v_\Delta - \mu \frac{v}{2} & \lambda_\Delta \frac{v_\Delta^2}{2} +  \mu \frac{v^2}{4v_\Delta} 
\end{pmatrix}\,.
\end{aligned}
\end{align}
The resulting mass eigenstates, in addition to  Im$(\phi^0)$, are
\begin{align}
\begin{aligned}
G^\pm &= \cos\zeta\ \phi^\pm + \sin\zeta\ \delta^\pm \,,\\
H^\pm &= - \sin\zeta\ \phi^\pm + \cos\zeta\ \delta^\pm \,,\\
h &= \cos\alpha\ {\rm Re} (\phi^0) + \sin\alpha\ \delta^0 \,,\\
H &= -\sin\alpha\ {\rm Re}(\phi^0) + \cos\alpha\ \delta^0\,,
\end{aligned}
\end{align}
with the CP-even and charged Higgs mixing angles
\begin{align}
\tan2\alpha &= \frac{4v v_\Delta \left(2\lambda_{\Phi\Delta} v_\Delta -\mu\right)}{2\lambda_\Phi v^2 v_\Delta -2\lambda_\Delta v_\Delta^3 -\mu v^2}\,, \;
\tan\zeta = -2 \frac{v_\Delta}{v}\,.
\end{align}
The massless states $G^\pm$ and Im$(\phi^0)$ are the {\it would-be} Goldstone bosons, eaten by the $W^{+}$ and $Z$. Among the massive states, $h$ is identified as the 125$\,$GeV (SM-like) Higgs, and $H$ and $H^\pm$ are the triplet-like neutral and charged scalars with masses
\begin{align}
\begin{aligned}
m_{H}^2 &= \lambda_\Delta \frac{v_\Delta^2}{2} +  \mu \frac{v^2}{4v_\Delta} - \tan\alpha \left(\lambda_{\Phi\Delta} v_\Delta -\frac{\mu}{2}\right) v\,,\\
m_{H^\pm}^2 &= \mu \frac{v^2+4v_\Delta^2}{4v_\Delta}\,.
\label{eq:mHp}
\end{aligned}
\end{align}
For vanishing $\alpha$, $m_{H^\pm}^2-m_{H}^2 \simeq \mu v_\Delta - \lambda_\Delta v_\Delta^2/2$, and thus the components are nearly mass-degenerate for $v_\Delta \ll v$. However, for large $\alpha$ and $v_\Delta$, vacuum stability and perturbative unitarity (see Sec. III of the main text) allow a mass-splitting $\Delta_m = m_{H^\pm}-m_H$ of a few GeV. {In addition, the EW radiative correction induces a mass-splitting of 160\,MeV--170\,MeV~\cite{Cirelli:2005uq}. However, such a small splitting is of little consequence as far as the LHC phenomenology and their contribution to the electroweak oblique parameters are concerned~\cite{Kanemura:2012rs,Cheng:2022hbo}.} 

Note that in the end, all parameters of the scalar potential can be expressed in terms of the physical masses and mixing angles and the two VEVs $v$ and $v_\Delta$. In particular, the (dimension-full) couplings of the neutral to the charged Higgses can be written as
\begin{align}
\begin{aligned}
A_{hH^\pm H^\mp} &\approx \frac{1}{2} \lambda_\Delta v_\Delta \sin\alpha + \lambda_{\Phi\Delta} v \cos\alpha\,,\\
A_{H H^\pm H^\mp} &\approx \frac{1}{2}\lambda_\Delta v_\Delta \cos\alpha - \lambda_{\Phi\Delta} v \sin\alpha\,,
\end{aligned}
\end{align}
in the limit of small $v_\Delta$ and $m_H\approx m_{H^\pm}$.

\section{\label{sec:stability_unitarity}Vacuum stability and perturbative unitarity}
In the following we provide the condition necessary to respect vacuum stability and perturbative unitarity (at tree level). The first can be derived requiring the potential to be bounded from below, while the latter, limiting the size of the quartic interactions, can be obtained from $2\to2$ scalar-scalar scattering\footnote{Note that this is valid as long as the $\mu$ parameter is not very large, which is satisfied for a small VEV since $\mu \sim v_\Delta \ll v$ for $m_{H^\pm} \sim v/2$.}.
The vacuum stability conditions read~\cite{Khan:2016sxm,Chabab:2018ert,FileviezPerez:2008bj}
\begin{align}
\label{eq:VS}
& \lambda_\Phi > 0, \quad \lambda_\Delta > 0, \quad \sqrt{2} \lambda_{\Phi\Delta} + \sqrt{\lambda_\Phi \lambda_\Delta} > 0,
\end{align}
and from perturbative unitarity we obtain
\begin{align}
\begin{aligned}
\label{eq:PU}
& |\lambda_\Phi| \leq 2\kappa \pi, \quad |\lambda_\Delta| \leq 2\kappa \pi, \quad |\lambda_{\Phi\Delta}| \leq \kappa \pi,
\\
& |6\lambda_\Phi + 5\lambda_\Delta \pm \sqrt{(6\lambda_\Phi - 5\lambda_\Delta)^2 + 192\lambda_{\Phi\Delta}^2}| \leq 8\kappa \pi,
\end{aligned}
\end{align}
where $\kappa$ = 16 or 8 depending on whether one chooses $|(a_0)| \leq 1$ or $|{\rm Re}~(a_0)| \leq \frac{1}{2}$ with $a_0$ denoting the leading partial wave amplitude~\cite{Logan:2022uus}. In order to be conservative, {\it i.e.~}not to exclude any potentially allowed parameter space, we opt for $\kappa=16$. Further, to ensure perturbativity at all higher orders, we require all the trilinear and quartic scalar couplings in \eqref{eq:pot} to be smaller than $4\pi$.\footnote{We checked numerically using {\tt Vevacious}~\cite{Camargo-Molina:2014pwa,Camargo-Molina:2013qva}, {\tt SPheno}~\cite{Porod:2003um,Porod:2011nf} and {\tt BSMArt}~\cite{Goodsell:2023iac} that the inclusion of the one-loop effective potential and meta stability has only a marginal effect on vacuum stability and perturbative unitarity.} 

\section{One loop functions for di-photon decay}

The loop functions for di-photon decay in Eq. (11) of the main text are given as
\begin{align}
\begin{aligned}
\beta^0_H(x) &= -x\left[1-xf(x)\right],
\\
\beta^{1/2}_H(x) &= 2x\left[1+(1-x)f(x)\right],
\\   
\beta^1_H(x) &= -\left[2+3x+3x(2-x)f(x)\right].
\end{aligned}
\end{align}  

The loop function for $H^\pm\to W H^0$ (in Eq. (20) of main text) is given by
\begin{align*}
& G(x,y) = \frac{1}{12y}\Bigg[2(-1+x)^3-9(-1+x^2)y+6(-1+x)y^2 
\\
& -6(1+x-y)y\sqrt{-\lambda(x,y)}\Bigg\{\tan^{-1}\left(\frac{1-x+y}{\sqrt{-\lambda(x,y)}}\right) 
\\
& +\tan^{-1}\left(\frac{1-x-y}{\sqrt{-\lambda(x,y)}}\right)\Bigg\}-3\left(1+(x-y)^2-2y\right)y\log x\Bigg],
\end{align*}
with $\lambda(x,y) = (1-x-y)^2 -4xy$.

\section{$H^\pm$ in stau searches}

\begin{figure*}[ht]
\centering
\includegraphics[scale=0.31]{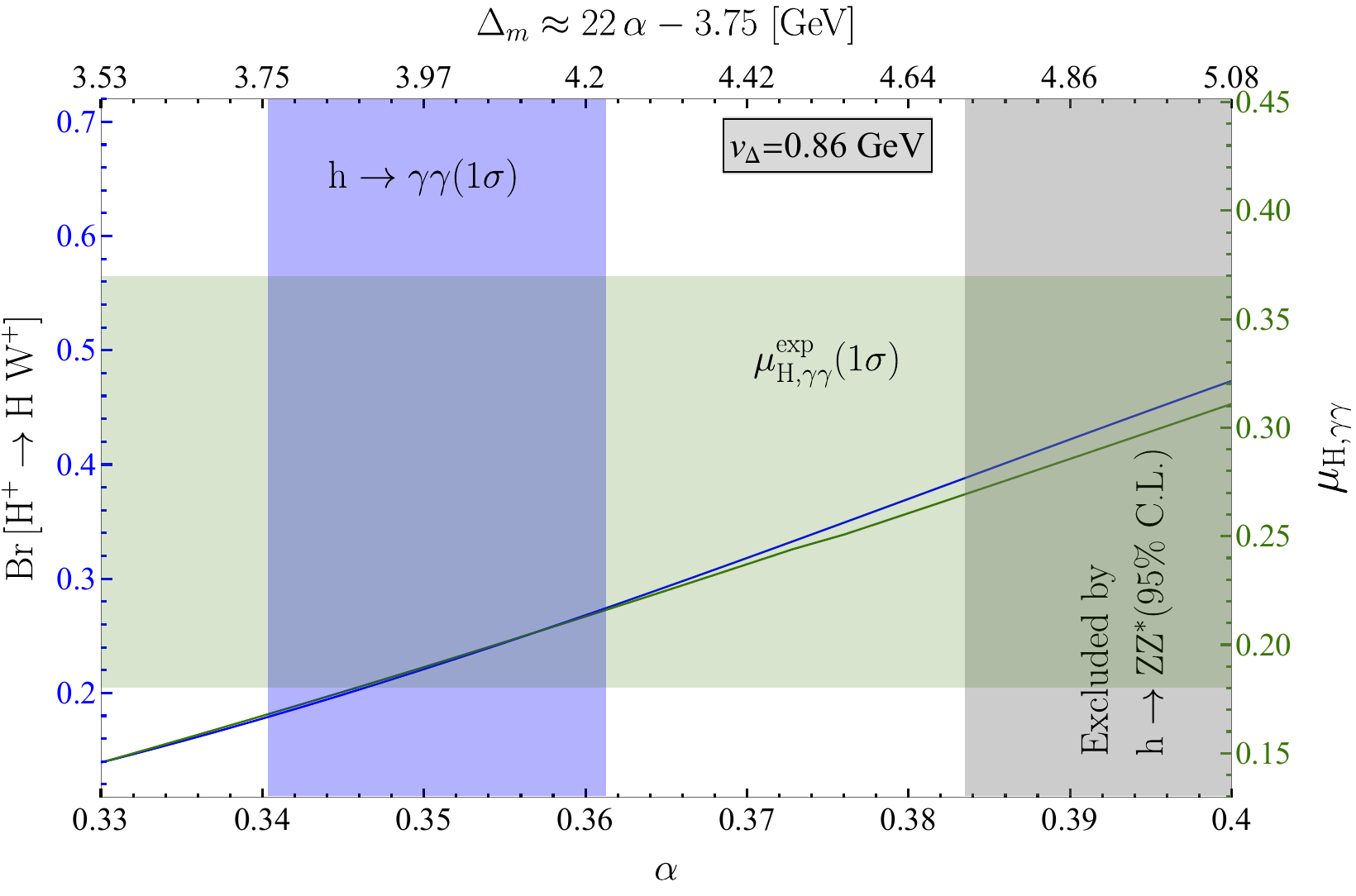}
\caption{Scenario with sizable $\Delta_m$ that is allowed by vacuum stability due to $\Delta_m \approx 22\alpha-3.75$ [GeV] that suppresses Br$[H^\pm\to\tau\nu]$ by enhancing Br$[H^\pm\to HW^*]$}
\label{masssplitting}
\end{figure*}

As to account for the slight tension with the 95\% exclusion limit of stau searches at the LHC, one can reduce Br$[H^\pm\to \tau\nu]$ by increasing the mass splitting $\Delta m$ such that Br$[H^\pm \to H W^{*}]$ becomes sizable. In Fig.~\ref{masssplitting}, we choose $v_\Delta= 0.86\,$GeV as a benchmark point, which allows for a small region in parameter space with sizable mass splitting, that is allowed by the vacuum stability and perturbative unitarity as well as compatible with $h\to\gamma\gamma,ZZ^*$, $H\to\gamma\gamma$ and $Zbb$. Note that this scenario predicts a small positive shift in the $W$ mass.

\newpage

\bibliographystyle{utphys}

\bibliography{apssamp}

\end{document}